\documentclass[%
 pra, twocolumn,
 amsmath,amssymb,
]{revtex4-2}

\UseRawInputEncoding

\usepackage{optidef}
\usepackage{braket}
\usepackage{nicematrix}
\usepackage{float}
\usepackage{multirow}
\usepackage{tikz}
\usepackage[utf8]{inputenc}

\usepackage{amsthm,color}

\usepackage{booktabs}
\usepackage{subcaption}
\usepackage{hyperref}
\usepackage{cleveref}

\usepackage{amsmath}
\usepackage{graphicx}
\usepackage{dcolumn}
\usepackage{bm}
\usepackage{ragged2e}

\begin{document}
\crefname{equation}{Eq}{Equations} 
\crefrangelabelformat{equation}{(#3#1#4--#5#2#6)}

\title{Time evolution of controlled many-body quantum systems with matrix product operators}

\author{
  Llorenç Balada Gaggioli$^{1,2}$  Jakub Mareček$^{1}$ \\[1em]
  \normalsize{$^{1}$ Czech Technical University in Prague, Prague, Czech Republic} \\
  \normalsize{$^{2}$ LAAS-CNRS, Université de Toulouse, France}
}

\date{\today}

\begin{abstract}
 We present a method for describing the time evolution of many-body controlled quantum systems using matrix product operators (MPOs). Existing techniques for solving the time-dependent Schrödinger equation (TDSE) with an MPO Hamiltonian often rely on time discretization. In contrast, our approach uses the Magnus expansion and Chebyshev polynomials to model the time evolution, and the MPO representation to efficiently encode the system's dynamics. This results in a scalable method that can be used efficiently for many-body controlled quantum systems. We apply this technique to quantum optimal control, specifically for a gate synthesis problem, demonstrating that it can be used for large-scale optimization problems that are otherwise impractical to formulate in a dense matrix representation.
\end{abstract}

\keywords{}
\maketitle
\allowdisplaybreaks

\section{Introduction}

The time-dependent Schrödinger equation (TDSE) is fundamental in quantum mechanics, as it governs the time evolution of quantum systems. Solving the TDSE accurately is crucial for the development of quantum technologies, understanding quantum dynamics and simulating physical processes. In fields such as quantum computing \cite{Ollitrault2021-fa}, quantum simulation \cite{Williams-Young2023-uw}, and quantum control \cite{Koch2022-al}, precise solutions to the TDSE are required for the manipulation and generation of quantum states and unitary transformations. Furthermore, solving the TDSE is crucial for the understanding of many-body quantum physics \cite{Nys2024-gf}, where the interactions between particles give rise to emergent phenomena such as entanglement or quantum phase transitions \cite{Hamazaki2021-gu}.

However, this is a challenging task due to the exponential growth of the Hilbert space with the system size. As a result, storing and manipulating state vectors and Hamiltonians becomes infeasible for large systems. If the system presents localized interactions, we can encode the information of the system's Hamiltonian in a tensor network like a matrix product operator (MPO)  \cite{chan2016matrixproductoperatorsmatrix, Parker_2020, PhysRevB.95.035129, Ba_uls_2023,ORUS2014117, Pirvu_2010, PhysRevA.81.062337, PhysRevB.78.035116, 10.1063/1.4939000}, which allows for a scalable representation of the system. 

Tensor networks have been used widely \cite{PhysRevB.88.085118,PhysRevB.90.045144,PhysRevB.90.115124,PhysRevB.105.165116, José_García-Ripoll_2006, Manmana_2005,Szalay_2015} in the study of the time evolution of quantum systems focusing mainly on the manipulation of high-dimensional quantum states. These methods make use of matrix product states (MPS) to efficiently represent the state and reduce computational complexity, allowing them to address multiple problems in condensed matter physics and quantum chemistry. In this work we extend this approach by developing a method that encodes the unitary evolution of a controlled quantum system in a tensor network, an MPO, which is useful for applications like quantum control, where the unitary evolution operator is essential for the design of gates in quantum computers.

The current uses of MPO Hamiltonians to model time evolution of many-body quantum systems often  rely on time-independent Hamiltonians in order to use methods like Trotter-Suzuki decomposition or commutator-free Magnus expansion \cite{Paeckel_2019, high-orderMPO,Miessen_2021,robertson2024tensornetworkenhanceddynamic,MPS_dynamics, Wall_2012}, which generally consider efficient state evolution instead of the unitary evolution. In this work we develop a fundamentally different method to solve the TDSE using tensor networks by representing the Hamiltonian as an MPO. Our method considers the continuous time interval and uses the Magnus expansion and Chebyshev polynomials to approximate the unitary operator that solves the TDSE in a scalable manner.

We apply our method to quantum optimal control problems, where finding the optimal control pulses to synthesize unitary gates or to prepare quantum states is crucial. Traditional quantum optimal control methods like GRAPE \cite{GRAPE} and CRAB \cite{CRAB} are also limited by the exponential scaling, although tensor network approaches have been explored to mitigate this issue \cite{CRAB_TN1,van_Frank_2016,CRAB_TN2,jensen2021approximatedynamicsleadoptimal}. By integrating the method developed in this paper with the quantum optimal control method QCPOP \cite{bondar2025globallyoptimalquantumcontrol,gaggioli2025unitarygatesynthesispolynomial}, which reformulates control problems as polynomial optimization problems, we achieve a scalable and efficient framework for solving quantum optimal control problems.

\section{Solution to the TDSE}

We want to solve the Schrödinger equation (where we let $\hbar=1$)
\begin{equation}
    \frac{\partial}{\partial t}U(t) = -iH(t) U(t), \quad U(0)=I.
    \label{schrodinger}
\end{equation}

We study a time-dependent Hamiltonian that is subject to a control function $u(t)$. Therefore, we have
\begin{equation}
    H(t)=H_0+u(t)H_c,
\end{equation}
where $H_0$ is the free Hamiltonian and $H_c$ the control Hamiltonian.

The solution to Equation \ref{schrodinger} can be written as the Magnus expansion \cite{Magnus}
\begin{equation}
    U(T)=\text{exp}(\Omega^{(\infty)}(T)),
\end{equation}
where
\begin{equation}
    \Omega^{(\infty)}(T) = \sum_{i=1}^{\infty}\Omega_i (T).
\end{equation}

If we let $H_j\equiv H(t_j)$ we can write the first terms of this sum as
\begin{align}
\label{Magnus1} \Omega_1(T) &= -i\int_0^T dt_1 H_1 \\
\label{Magnus2}\Omega_2(T) &= \frac{-1}{2} \int_0^T  dt_1  \int_0^{t_1}  dt_2 [H_1,H_2] \\
\label{Magnus3}\Omega_3(T) &= \frac{i}{6} \int_0^T  dt_1  \int_0^{t_1}  dt_2  \int_0^{t_2}  dt_3 \bigg(\Big[H_1,[H_2,H_3]\Big]+\\
& \hspace{40pt}+\Big[[H_1,H_2],H_3\Big] \bigg) \notag.
\end{align}
The nested commutators enclose the time-ordering effects of the system, and absolute convergence is assured \cite{Blanes_2009} if 
\begin{equation}
    \int_0^{T}\|H(t)\|_2 dt <\pi.
    \label{convergence}
\end{equation}

Furthermore, we can approximate the exponential of the Magnus series truncated to a certain order, $n$, through a Chebyshev polynomial \cite{Chebyshev} in the following way
\begin{equation}
    U(T)\approx\text{exp}(\Omega^{(n)})\approx J_0(1)\mathbb{I}+2\sum_{i=1}^{p}J_i(1)T_i,
\end{equation}
where $p$ is the truncation order, $J_i(x)$ is the Bessel function, and $T_i$ is the Chebyshev polynomial element defined recursively as
\begin{equation}
    T_0=\mathbb{I}, \quad T_1=\Omega^{(n)}, \quad T_{i+1}=2\Omega^{(n)}T_i+T_{i-1}.
\end{equation}

\section{Matrix product operators}

In this section we will go through the main concepts underlying matrix product operators to see how their structure changes when we do operations with them. For a more detailed introduction to tensor networks, and MPOs specifically, we recommend \cite{ORUS2014117, chan2016matrixproductoperatorsmatrix,Parker_2020}.

If we assume the physical system we are controlling is an $N$-body system and has some type of localized interactions, we can write the Hamiltonian terms as MPOs in an efficient manner. By localized interactions we mean that each site interacts only with neighboring sites, typically those adjacent or within a short range. This characteristic leads to sparse operator structure, with low entanglement, which results in a small MPO bond dimension, which makes the MPO representation efficient. We let

\begin{equation}
    H_0=\sum_{\mathbf{s},\mathbf{s'}}W^{[1]}_{s_1 s'_1}\hdots W^{[N]}_{s_N s'_N} \ket{\mathbf{s}}\bra{\mathbf{s'}},
\end{equation}
where $\ket{\mathbf{s}}\equiv\ket{s_1s_2\hdots s_N}$. For $i\in\{2,\hdots,N-1\}$, $W^{[i]}$ are order 4 tensors with dimensions $d_i\times d'_i\times r_i\times r'_i$. $W^{[1]}$ has dimensions $d_1\times d'_1\times 1\times r_i$ and $W^{[N]}$ has dimensions $d_{N}\times d'_{N}\times r_{N}\times 1$, both of which impose the boundary conditions. 

We will work with qubits so we let the physical dimensions be $d_i=d'_i=2$ for all tensors, which implies $s_i,s'_i\in \{0,1\}$ for all $i$. The artificial bond dimensions $r_i,r'_i$ represent the interactions between sites and it can change depending on the system.

We can also see the tensors as matrices $W^{[2]}_{s_2 s'_2},\hdots,W^{[N-1]}_{s_{N-1} s'_{N-1}}$ of size $r_i\times r'_i$, a row vector $W^{[1]}_{s_1 s'_1}$ of size $1\times r_1$ and a column vector $W^{[N]}_{s_N s'_N}$ of size $r_N\times 1$. We illustrate the MPO decomposition in Figure ~\ref{fig:MPO}, where we show at a high-level how we reduce the number of elements, from $O(4^N)$ to $O(N)$ if we fix the physical and bond dimensions, needed to describe the Hamiltonian by finding local tensors, $W^{[i]}$, for each site. The sizes of $s_i,s'_i$ are fixed as they are physical quantities, but the size of the bond dimensions, $r_i$, can vary; and consequently, the accuracy of the MPO representation also varies.

\tikzstyle{bigmatrix}=[rectangle,draw=red!60,fill=red!20,thick,minimum width=1.8cm, minimum height=1.8cm]
\tikzstyle{tensor}=[rectangle,draw=blue!50,fill=blue!20,thick,minimum width=1cm, minimum height=1cm]
\tikzstyle{line}=[thick]

\begin{figure}[h]
    \centering
    \resizebox{\columnwidth}{!}{ 
    \begin{tikzpicture}[inner sep=0.6mm, node distance=1.2cm]

        \node[bigmatrix] (H) at (-3.2,0) {$H$};
        \node at (-2.2, 1.2) {\footnotesize $\sim O(4^N)$};

        \node (s1H) at (-3.2, -1.5) {\footnotesize $\mathbf{s' }$};
        \node (s2H) at (-3.2, 1.5) {\footnotesize $\mathbf{s}$};
        \draw[-, line] (H) -- (s1H);
        \draw[-, line] (H) -- (s2H);
        
        \draw[->, line] (-2,0) -- (-0.2,0) node at (-1.1,0.25) {\footnotesize MPO};

        \node[tensor] (W1) at (1, 0) {\footnotesize $W^{[1]}$};
        \node[tensor] (W2) at (2.8, 0) {\footnotesize $W^{[2]}$};

        \node at (4.2, 0) {\footnotesize $\dots$};

        \node[tensor] (WN) at (5.6, 0) {\footnotesize $W^{[N]}$};

        \draw[-, line] (W1.east) -- (W2.west) node[midway, above] {\footnotesize $r_1$};
        \draw[-, line] (W2.east) -- (4,0) node[midway, above] {\footnotesize $r_2$}; 
        \draw[-, line] (4.4,0) -- (WN.west) node[midway, above] {\footnotesize $r_N$};

        \foreach \i/\x in {1/1, 2/2.8, N/5.6} {
            \node (s\i) at (\x, -1.2) {\footnotesize $s'_\i$};  
            \node (s\i up) at (\x, 1.2) {\footnotesize $s_\i$};  
            \draw[-, line] (W\i) -- (s\i);  
            \draw[-, line] (W\i) -- (s\i up);  
        };

        \node at (4.2, 1) {\footnotesize $\sim O(N)$};

    \end{tikzpicture}
    } 
    \caption{\justifying MPO decomposition of an $N$-qubit Hamiltonian.}
    \label{fig:MPO}
\end{figure}
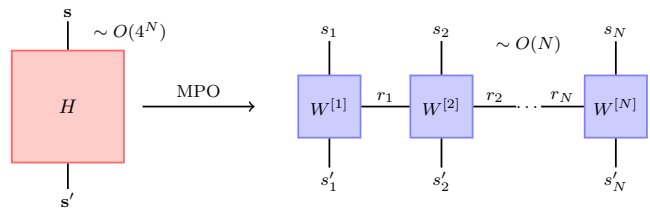

If we let the bond dimension for all tensors to be $r=3$, then we can visualize the central tensors as
\begin{equation}
    W^{[i]}_{s_i s'_i}=\begin{pmatrix}
        \begin{pmatrix}
            * & * \\ * & *
        \end{pmatrix}_{s_i s'_i} & \begin{pmatrix}
            * & * \\ * & *
        \end{pmatrix}_{s_i s'_i} & \begin{pmatrix}
            * & * \\ * & *
        \end{pmatrix}_{s_i s'_i}\\
        \begin{pmatrix}
            * & * \\ * & *
        \end{pmatrix}_{s_i s'_i} & \begin{pmatrix}
            * & * \\ * & *
        \end{pmatrix}_{s_i s'_i} & \begin{pmatrix}
            * & * \\ * & *
        \end{pmatrix}_{s_i s'_i}\\
        \begin{pmatrix}
            * & * \\ * & *
        \end{pmatrix}_{s_i s'_i} & \begin{pmatrix}
            * & * \\ * & *
        \end{pmatrix}_{s_i s'_i} & \begin{pmatrix}
            * & * \\ * & *
        \end{pmatrix}_{s_i s'_i}
    \end{pmatrix}\notag.
\end{equation}

The initial and final tensors follow similarly but in the shape of row and column vectors.

We also let
\begin{equation}
    H_c=\sum_{\mathbf{s},\mathbf{s'}}V^{[1]}_{s_1 s'_1}\hdots V^{[N]}_{s_N s'_N}\ket{\mathbf{s}}\bra{\mathbf{s'}}.
\end{equation}

Let us now look at how to perform operations with MPOs. We consider scalar products, sums and products, which are all the necessary ingredients to do the Magnus expansion and the Chebyshev approximation, as we show in Appendix \ref{tensors}.

The scalar product only affects one of the tensors, the first one for example, so for a scalar $a$ we write
\begin{equation}
    aH_0=\sum_{\mathbf{s},\mathbf{s'}}aW^{[1]}_{s_1 s'_1}\hdots W^{[N]}_{s_N s'_N} \ket{\mathbf{s}}\bra{\mathbf{s'}}.
\end{equation}

For the sum of two MPOs we write
\begin{align}
    H_0+H_c&=\sum_{\mathbf{s},\mathbf{s'}}W^{[1]}_{s_1 s'_1}\hdots W^{[N]}_{s_N s'_N} \ket{\mathbf{s}}\bra{\mathbf{s'}}\notag\\
    &\hspace{40pt}+\sum_{\mathbf{s},\mathbf{s'}}V^{[1]}_{s_1 s'_1}\hdots V^{[N]}_{s_N s'_N} \ket{\mathbf{s}}\bra{\mathbf{s'}}\notag\\
    &=\sum_{\mathbf{s},\mathbf{s'}}\Big(W^{[1]}_{s_1 s'_1}\oplus V^{[1]}_{s_1 s'_1} \Big)\hdots \notag\\
    &\hspace{40pt}\hdots\Big(W^{[N]}_{s_N s'_N}\oplus V^{[N]}_{s_N s'_N}\Big) \ket{\mathbf{s}}\bra{\mathbf{s'}}.
\end{align}

And for the product we have
\begin{align}
    H_0H_c&=\sum_{\mathbf{s},\mathbf{s'}}W^{[1]}_{s_1 s'_1}\hdots W^{[N]}_{s_N s'_N} \ket{\mathbf{s}}\bra{\mathbf{s'}}\notag\\
    &\hspace{40pt} \sum_{\mathbf{s},\mathbf{s'}}V^{[1]}_{s_1 s'_1}\hdots V^{[N]}_{s_N s'_N} \ket{\mathbf{s}}\bra{\mathbf{s'}}\notag\\
    &=\sum_{\mathbf{s},\mathbf{s'}}\Big(\sum_{l_1}W^{[1]}_{s_1 l_1}\otimes V^{[1]}_{l_1 s'_1} \Big)\hdots\notag\\
    &\hspace{40pt}\hdots\Big(\sum_{l_N}W^{[N]}_{s_N l_N}\otimes V^{[N]}_{l_N s'_N}\Big) \ket{\mathbf{s}}\bra{\mathbf{s'}}.
\end{align}

Using these operations we can write the solution to the TDSE in an MPO form
\begin{align}
    U(T)=\sum_{\mathbf{s},\mathbf{s'}}T^{[1]}_{s_1 s'_1}\hdots T^{[N]}_{s_N s'_N}\ket{\mathbf{s}}\bra{\mathbf{s'}}.
\end{align}

We note that, by taking sums and products, the bond dimension of the resulting MPO $U(T)$ increases. In Table \ref{tab:bond_dimensions} we see how the bond dimension of $U(T)$ increases with the higher truncation orders we make in the Magnus and Chebyshev approximation.

\begin{table}[h]
    \centering
    \renewcommand{\arraystretch}{1.4}
    \setlength{\tabcolsep}{4pt}
    \begin{tabular}{ccccc}
        \toprule
        \multirow{2}{*}{\textbf{Magnus Order}} & \multicolumn{4}{c}{\textbf{Chebyshev Order}} \\
        \cmidrule(lr){2-5}
         & 1st & 2nd & 3rd & 4th \\
        \midrule
        1st  & $7$ & $43$ & $259$ & $1.55\cdot 10^{3}$ \\
        2nd  & $25$ & $601$ & $1.44\cdot10^4$ & $3.46\cdot 10^5$ \\
        3rd  & $187$ & $3.47\cdot 10^4$ & $6.46\cdot 10^6$ & $1.20\cdot 10^9$ \\
        4th  & $1.15\cdot 10^{3}$ & $1.34\cdot 10^6$ & $1.55\cdot 10^9$ & $1.79\cdot 10^{12}$ \\
        \bottomrule
    \end{tabular}
    \caption{\justifying Bond dimensions for different orders of the Magnus expansion and Chebyshev approximation, for initial bond dimension of 3 for both $H_0$ and $H_c$.}
    \label{tab:bond_dimensions}
\end{table}

From Table \ref{tab:bond_dimensions} we see that to keep the bond dimension of the tensors around $10^{4}$, which would be enough to do computations with a laptop, we can use a Magnus expansion of order 2 and Chebyshev approximation of order 3 or vice-versa. However, the characteristics of the physical system we are considering will be the main indicator to know the order at which we should truncate the approximations. 

If we use a dense matrix method, where we operate with matrices of $4^N$ elements, we reach a computational limit relatively fast. Our  method, on the other hand, uses MPOs storing $4Nr^2$ elements, which scales linearly in $N$ instead of exponentially. This allows for the use of the MPO method for very large systems.

As we see, the biggest limitation of this method is the growth of the bond dimension, but this is a recurring theme when working with MPOs, as there is no physically-sensitive method to reduce the bond dimension of the unitary operator written as an MPO \cite{high-orderMPO}, in the same sense as we can use bond dimension reduction algorithms for MPS. Currently existing methods \cite{Paeckel_2019, high-orderMPO,Miessen_2021,robertson2024tensornetworkenhanceddynamic,MPS_dynamics, Wall_2012}  are based on building the unitary operator as an MPO for each time step, then applying it to the MPS, and finally reducing the bond dimension to make the full time evolution computationally feasible. We can also consider the time evolution of the unitary operator and limit the number of time steps according to the maximum bond dimension we can have for the resulting MPO. Our method lies on this line as it does not (and it cannot) depend on bond dimension reduction algorithms. Firstly, because we work with operators, not states, and secondly, because this allows us to work with variables symbolically in the tensor network entries, which will help us formulate the quantum optimal control problem. The bond dimension reduction algorithms are numerical and therefore cannot work symbolically.

\section{Example}
Let us look at the controlled Ising model, for which we have
\begin{equation}
    H(t)=\sum_{i=1}^{N-1} J \sigma_i^z \sigma_{i+1}^z + u(t)\sum_{i=1}^N\sigma_i^x.
\end{equation}

To ensure convergence of the Magnus expansion through Equation \ref{convergence} we will set the maximum time evolution for this Hamiltonian to be $T_{\text{max}}=\frac{1}{2N}$, considering its norm scales with $N$. We take the control function to be the polynomial $u(t)=1-t+t^2$ and $J=1$. Note that $u(t)$ can take any form as long as it is integrable.

We write the free Hamiltonian $H_0$ as an MPO with boundary tensors
\begin{equation}
    W^{[1]} = \begin{pmatrix}
        0 & J\sigma^{z} & I
    \end{pmatrix}, \quad W^{[N]}=\begin{pmatrix}
       I \\ \sigma^{z}\\ 0
    \end{pmatrix},
    \label{MPO1}
\end{equation}

and central tensors, for $i=2,\hdots,N-1$, of the form
\begin{equation}
    W^{[i]} = \begin{pmatrix}
        I & 0 & 0\\
        \sigma^z & 0 & 0\\
        0 & J\sigma^{z} & I
    \end{pmatrix}.
    \label{MPO2}
\end{equation}

Similarly, for the control Hamiltonian $H_c$ we have boundary tensors
\begin{equation}
    W^{[1]} = \begin{pmatrix}
        I & \sigma^{x}
    \end{pmatrix}, \quad W^{[N]}=\begin{pmatrix}
        \sigma^{x}\\ I
    \end{pmatrix},
    \label{MPO3}
\end{equation}
and core tensors, for $i=2,\hdots,N-1$, of the form
\begin{equation}
    W^{[i]} = \begin{pmatrix}
        I & \sigma^{x}\\
        0 & I
    \end{pmatrix}.
    \label{MPO4}
\end{equation}

We note that the initial bond dimensions are 3 and 2 so the total bond dimension will increase at a slightly slower pace than Table \ref{tab:bond_dimensions}. We now look at the error between the dense representation ($2^N\times 2^N$ matrix) of the unitary and the one resulting from the Magnus and Chebyshev approximation using MPOs.

To look at the error $|\varepsilon|$ between the dense unitary, $U_{\text{dense}}$, and the approximate unitary in MPO form, $U_{\text{MPO}}$, we compute the infidelity between them, defined as
\begin{equation}
    \varepsilon = 1 - \frac{1}{d^2}|\text{Tr}(U_{\text{dense}}^{\dag} U_{\text{MPO}})|^2,
\end{equation}

where $d$ is the dimension of the system. We consider the infidelity instead of other error measures because we are interested in the gate synthesis problem, and the infidelity is directly related to the error probabilities we see in experiments.

\begin{figure}[h]
    \centering

    \includegraphics[width=1\linewidth]{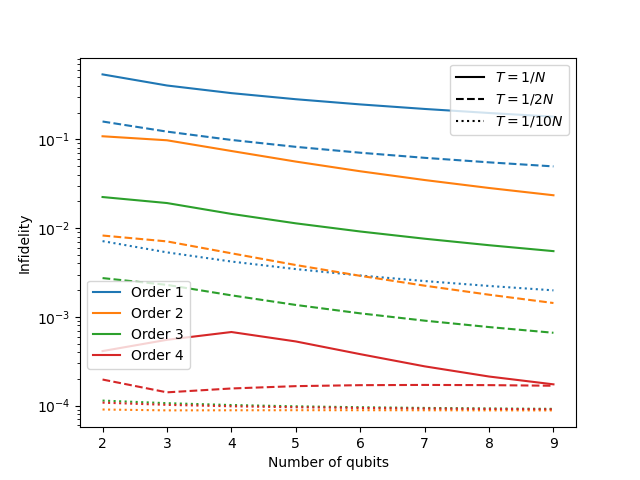}

    \caption{\justifying Infidelity between the dense and the MPO method. We fix a Magnus expansion of order 1 and vary the Chebyshev order. We compare different time evolutions.}
    \label{fig:infidelity_change}
\end{figure}

The error results only from the approximations and it is independent of the use of MPO representations which, in this case, is exact. We plot the error for different orders of truncation of the Chebyshev approximation and for different time evolutions in Figure ~\ref{fig:infidelity_change}.

We only consider Magnus expansions of order 1 because, for this Hamiltonian, higher orders do not improve the accuracy. From Figure ~\ref{fig:infidelity_change}, we note that to get an error of less than $1\%$, for $T=\frac{1}{N}$, we need a Chebyshev approximation of, at least, order 3. Similarly, we see how the shorter the evolution time, the better the infidelity. From these, we decide the truncation orders of the unitary calculation for this example to be 1 for the Magnus expansion and 3 for the Chebyshev approximation. We will let the time evolution be $T=\frac{1}{2N}$.

\begin{figure}[h]
    \centering
    \includegraphics[width=1\linewidth]{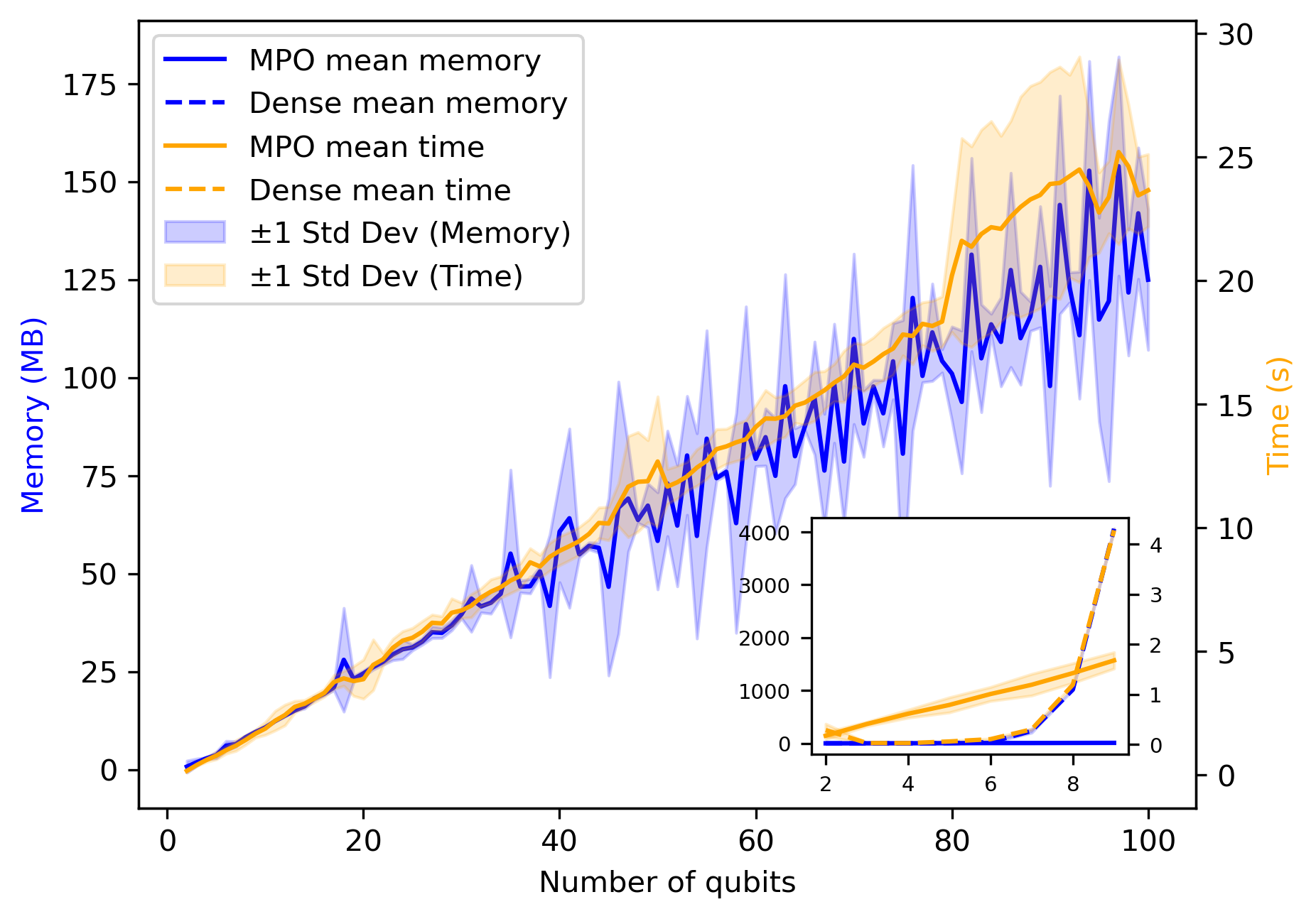}
    \caption{\justifying Time and memory required to compute the dense and MPO unitary evolutions using a first order Magnus expansion, a Chebyshev approximation of third order, and an evolution time $T=\frac{1}{2N}$. We do 5 repetitions for each $N$ to look at the mean and standard deviation.}
    \label{fig:Times}
\end{figure}

To see how the MPO representation provides an advantage compared to the full matrix representation we look at the time and memory required to compute both the dense and MPO solutions, and plot these in Figure ~\ref{fig:Times}.

We use the library \textit{QuTiP} \cite{qutip} in Python to calculate the dense evolution, which can be solved almost exactly. In Figure ~\ref{fig:Times}, we see that for a small number of qubits ($N<8$) it takes less resources to compute the evolution using a dense representation. However, as the number of qubits increases, the resources required for the unitary evolution increase exponentially, making the computation infeasible for large $N$. On the other hand, the resources required for the MPO method only increase linearly. Therefore, for more than 8 qubits we have an advantage by using the MPO representation.

We can also compare our method with a Trotterization method, where we build an MPO following a discrete time approach, dividing the time interval into $K$ segments such that $T=\Delta t K$. Therefore, we write the total time evolution of the unitary operator as $U(T)=U_K U_{K-1}\hdots U_2 U_1$, where $U_j=e^{-i\Delta t H(\Delta t j)}$. If we consider the controlled Ising model we can approximate this by considering almost-commuting parts of the Hamiltonian 

\begin{align}
H(t)&=H_Z^{odd}+H_Z^{even} + H_X(t)\\
&=\sum_{i \text{ odd}}^{N-1} J \sigma_i^z \sigma_{i+1}^z + \sum_{i \text{ even}}^{N-1} J \sigma_i^z \sigma_{i+1}^z + u(t)\sum_{i=1}^N\sigma_i^x
\end{align}

such that $U_j \approx e^{-i\Delta t H_X(\Delta t j)} e^{-i\Delta t H_Z^{odd}} e^{-i\Delta t H_Z^{even}}$, with an error of $O(\Delta t ^2)$.

The terms of the sums of these 3 components all commute with themselves, which allows us to write 

\begin{align}
    e^{-i\Delta t H_X(\Delta t j)}&=\prod_{i}^N e^{-i\Delta t u(\Delta t j) \sigma_i^x}\\
    e^{-i\Delta t H_Z^{odd}}&=\prod_{i \text{ odd}}^{N-1} e^{-i\Delta t J\sigma_i^z \sigma_{i+1}^z}\\
    e^{-i\Delta t H_Z^{even}}&=\prod_{i \text{ even}}^{N-1} e^{-i\Delta t J\sigma_i^z \sigma_{i+1}^z}.
\end{align}
These factorizations allow for simple MPO representations, more details in Appendix ~\ref{trotterization}. The MPOs have bond dimensions of $[1,2,2]$ respectively, which means one time step of the unitary operator will be an MPO of bond dimension 4. The total bond dimension for the MPO of $U(T)$ will then be $4^K$, and if we want to keep the bond dimension below computational limits, such as below a bond dimension of $\sim 10^4$, then we can have a maximum of 7 time steps per interval. 

\begin{figure}[h]
    \centering
    \includegraphics[width=1\linewidth]{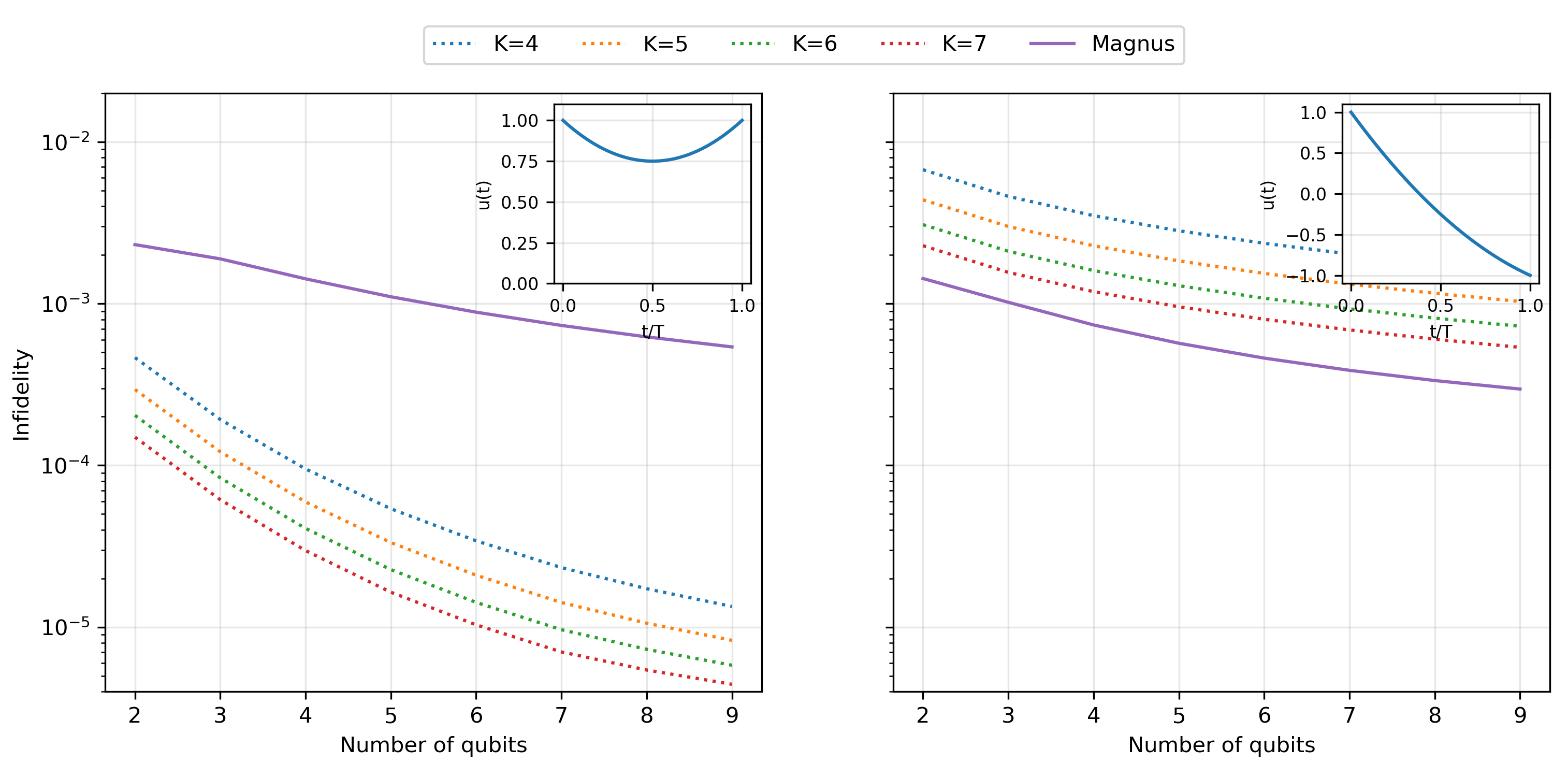}
    \caption{\justifying Comparison of accuracy of the Magnus and Chebyshev method with Trotterization of $K$ time steps for two different controls $u(t)$.}
    \label{fig:trotter}
\end{figure}

We compare the accuracy of this Trotterization of first order with our method of Magnus of first order and Chebyshev approximation of third order, and illustrate the results in Figure ~\ref{fig:trotter}. We let $T=\frac{1}{2N}$, $J=1$, and compare two different control functions. Note that, the bond dimension of the Trotter method will be $\sim 10^2$ times bigger.

Trotterization provides better accuracy for controls with not much steepness or with oscillations, while the Magnus and Chebyshev perform better for steeper controls that Trotterization might undersample due to the low number of time steps. This indicates that both approaches can still be used to find the unitary evolution of systems that can be represented with low bond dimensions, and depending on the characteristics of the system we might prefer one or the other. In the next section we show how for an application like quantum optimal control of many body systems, our approach might provide even a larger advantage.

The limitations of our method are illustrated by the results in Table ~\ref{tab:bond_dimensions} and Figure ~\ref{fig:infidelity_change}, in the former we see the bond dimension increases very fast as we improve the order of approximations, which means that for Hamiltonians with higher starting bond dimension (or equivalently, longer-range interactions) we might not be able to have an accurate approximate solution to the unitary operator. In the latter, we consider the other limiting factor, the total time evolution $T$, which has to be very small in order for the Magnus expansion to fulfill its convergence properties. In what follows, we will improve the fast bond dimension growth by avoiding the Chebyshev polynomial approximation, leading to accurate results for quantum optimal control.

\section{Quantum optimal control}

We now consider a quantum optimal control problem following from the previous example to show the power of the MPO representation method. We take the Hamiltonian of the system to be 
\begin{equation}
    H(t,\mathbf{x})=\sum_{i=1}^{N-1} J \sigma_i^z \sigma_{i+1}^z + u(t,\mathbf{x})\sum_{i=1}^N\sigma_i^x,
\end{equation}

with $J=1$ and a control function 
\begin{equation}
    u(t,\mathbf{x})=\sum_{i=1}^m x_i t^{i-1},
\end{equation}
where $\mathbf{x}=\{x_1,\hdots , x_m\}$ are unknown control coefficients we want to optimize. Now we represent the Hamiltonian terms as MPOs using the tensors defined in \cref{MPO1,MPO2,MPO3,MPO4}.

We want to synthesize a unitary gate, for example the global rotation
\begin{equation}
    U^{\star} = e^{-i\frac{\pi}{4}\sum_{i=1}^{N-1} \sigma^z_i \sigma^z_{i+1}},
\end{equation}
which is a multi-qubit controlled-Z gate (CZ).

There are several requirements the target unitary must have for a good formulation of the problem. The most important one is that we can write it as the exponential of an MPO, even for large bond dimension $r$. Secondly, if we want to synthesize the specific target, and not just minimize the error, we require the unitary target to be reachable for our defined system. This means that there exists a control function, $u(t)$, and finite time, $T$, such that the Schrödinger equation evolves the system from the identity to $U_{\text{target}}$. This restricts the choice, and that is why we pick $U^{\star}$ such that we can reach it and write it as the exponential of an MPO.

In order to optimize the control parameters of $u(t,\mathbf{x})$ to reach the desired unitary gate we have to solve the TDSE with the defined control Hamiltonian. If we pick a final evolution time $T$ we can find the solution $U(T,\textbf{x})$ using the Magnus and Chebyshev approximations.

Once we have the evolution solution in terms of the unknown control coefficients $\textbf{x}$ we want to find the optimal values for which we get as close as possible to the desired unitary gate. To find these we formulate the optimization problem 
\begin{equation}
    \min_{\mathbf{x}} \quad 1 - \frac{1}{d^2}|\text{Tr}(U^{\star\dag}U(T,\mathbf{x}))|^2,
    \label{optim}
\end{equation}
where $d$ is the dimension of the Hilbert space.

We want to find the global solution and recover the optimal control values that minimize this function. For this we use QCPOP, the method shown in \cite{bondar2025globallyoptimalquantumcontrol, gaggioli2025unitarygatesynthesispolynomial} with which we can reformulate problem \ref{optim} as a polynomial optimization problem of the following form
\begin{equation}
    \min_{\mathbf{x}} \quad \|\Omega^{(n)}(T,\mathbf{x})-i\Theta\|_F^2,
    \label{problem}
\end{equation}
where $\Omega^{(n)}(T,\mathbf{x})$ is the truncated Magnus expansion at order $n$, and $\Theta$ is the Hermitian matrix generating the unitary target, in our case $\Theta=-\frac{\pi}{4}\sum_{i=1}^{N-1} \sigma^z_i \sigma^z_{i+1}$. 

Now we assume that $\Theta$ can be efficiently represented as an MPO and therefore we can use the MPO operations defined previously to reformulate \ref{problem} as
\begin{align}
    &\min_{\mathbf{x}} \quad \text{Tr}\bigg((-\Omega^{(n)}(T,\mathbf{x})+i\Theta)(\Omega^{(n)}(T,\mathbf{x})-i\Theta)\bigg).
\end{align}

This is a polynomial that can then be minimized using the moment-SOS method \cite{lasserre}, a series of relaxations of semidefinite programs that will generally converge to the global solution. To implement this numerically, and letting $m=3$, we use the Julia package \textit{TSSOS} \cite{TSSOS} in order to globally solve the polynomial optimization problem and extract the solution of the control parameters. In this example we find $\mathbf{x}=[0,0,0]$.

In QCPOP we also use the Magnus expansion and Chebyshev approximation, so the error in the approximation arises from the Magnus expansion only and not from the MPO representation. Therefore, we will obtain the same polynomial to minimize using both the MPO and dense method. The only difference between both methods lies in the time and memory requirements to obtain the polynomial, and here is where we have the advantage of using the MPO representation. 

\begin{figure}[H]
    \centering
    \includegraphics[width=1\linewidth]{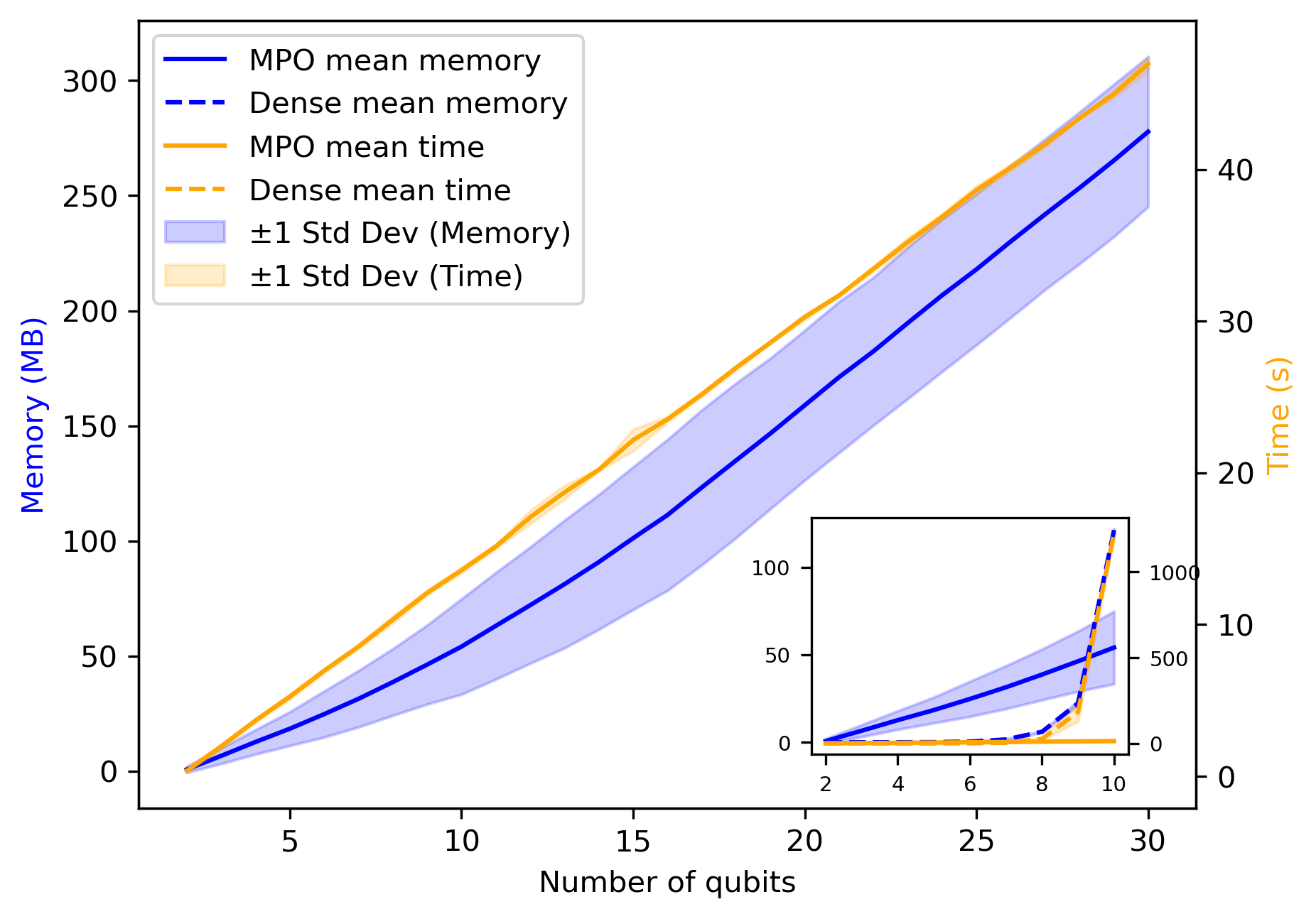}
    \caption{\justifying Time and memory required to compute the polynomial to minimize for the dense and MPO representations. We use a second order Magnus expansion, an evolution time $T=\frac{\pi}{4}$, and a control function with 3 unknown parameters. We do 5 repetitions per number of qubits to look at the mean and standard deviation.}
    \label{fig:Time_mem}
\end{figure}

In Figure ~\ref{fig:Time_mem}, we see how, for a control polynomial with $m=3$ unknown parameters, the time and memory required to compute the polynomial to optimize are linear for the MPO representation, and exponential for the dense representation. 

Now, we compare the error of the QCPOP method with the errors of the currently used methods GRAPE \cite{GRAPE} and CRAB \cite{CRAB}, with the initial guess as a random control function. The polynomial we obtain from our tensor network approach is the same as the dense matrix method, and the tensor network applications in GRAPE and CRAB are formulated for MPS evolution. Therefore, we do an error comparison without using tensor networks, which will only make the accuracy of GRAPE and CRAB worse, while QCPOP will remain the same.

\begin{figure}[H]
    \centering
    \includegraphics[width=1\linewidth]{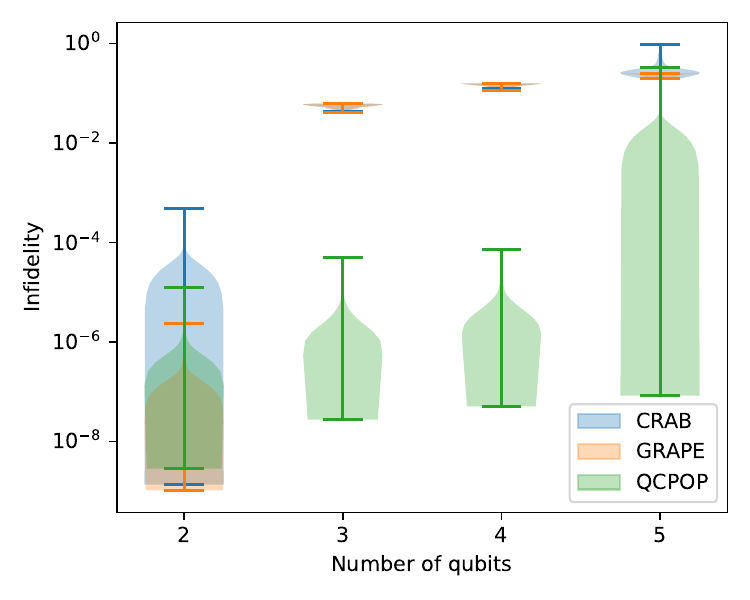}
    \caption{\justifying Infidelity comparison between different quantum optimal control methods for the controlled Ising model with different number of qubits, for 100 random reachable unitary targets. The color shading is the range of infidelities and the width is the frequency. For 3 and 4 qubits, CRAB and GRAPE almost fully intersect, making them hard to distinguish.}
    \label{fig:qoc_compare}
\end{figure}

In particular, we sample from an ensemble of random control parameters $\mathbf{x}=\{x_1,x_2,x_3\}$ for which $x_1,x_2,x_3\in [-1,1]$. For each control signal, we simulate the evolution using \textit{QuTiP} \cite{qutip} to obtain the target unitary. Subsequently, we solve the quantum optimal control problem for the target unitary with the first level of the moment-SOS hierarchy and a second order Magnus expansion. We do this for 100 different control parameters, and for different number of qubits in the Ising Hamiltonian model. We illustrate the results in Figure ~\ref{fig:qoc_compare}.

For very small system size (2 qubits), we see that GRAPE performs the best followed closely by QCPOP and then CRAB. As we increase the system size, all three methods get worse but, while GRAPE and CRAB become unreliable methods with high infidelities, QCPOP generally maintains the infidelity below $1\%$, a major improvement with respect to the other methods.

\section{Conclusion}

We have presented a method to represent the time evolution of controlled quantum systems using matrix product operators. This allows for the calculation of the approximate unitary operator solution to the TDSE of large quantum systems in a scalable manner. We improve the exponential matrix size increase in dense matrix methods by using matrix product operators, which scale linearly in size. These scaling properties can be used to solve quantum optimal control problems for large quantum systems, which would be infeasible to calculate using a dense matrix form. Our use of the MPOs is limited to systems that present specific properties related to localized interactions, which allow us to represent them efficiently using tensor networks, and for short time evolution. Using the MPO representation there is no direct loss in accuracy, and the quality of the approximation is fully determined by the orders of truncation of the Magnus and Chebyshev approximations. The lack of loss in accuracy of the MPO solution with respect to the dense matrix solution arises from not using bond reduction algorithms. This allows us to both work with unitary operators without considering the state evolution, and also using unknown coefficients in the tensor network, which leads to the formulation of the quantum optimal control problem of gate synthesis, through QCPOP, that we show improves upon current methods for the example provided.

\section*{Acknowledgments}
This work has been supported by European Union’s HORIZON–MSCA-2023-DN-JD programme under under the Horizon Europe (HORIZON) Marie Skłodowska-Curie Actions, grant agreement 101120296 (TENORS). J.M. has been supported by the Czech Science Foundation (23-07947S).

\section*{Data availability}

All codes used in this work can be found in \textit{https://github.com/llorebaga/TEMPO}

\appendix

\section{Operations to solve the TDSE}
\label{tensors}

We want to show how the only necessary operations to compute $U(T)$ are scalar products, sums and products of MPOs. For this, let us write the terms of the Magnus expansion
\begin{align}
    \Omega_1 &= -i\int^T_0  H(t)dt\notag\\
    &= -i\int^T_0  H_0 + u(t)H_c dt\notag\\
    & = \Big(-i\int^T_0 dt\Big)H_0+\Big(-i\int^T_0 u(t)dt\Big)H_c\notag\\
    & = aH_0+bH_c,
\end{align}
for scalars $a,b$. From now on we use the notation $H(t_i)=H_i$ and $u(t_i)=u_i$ Similarly, the second term will be
\begin{align}
    \Omega_2 &= -\frac{1}{2}\int^T_0 dt_1\int^{t_1}_0 [H_1,H_2]dt_2\notag\\
    &=\Big( -\frac{1}{2}\int^T_0 dt_1\int^{t_1}_0(u_2-u_1)dt_2\Big)[H_0,H_c]\notag\\
    &= c [H_0,H_c],
\end{align}
for scalar $c$. For the third term we have the following
\begin{align}
    &\Omega_3(T) = \frac{i}{6} \int_0^T  dt_1  \int_0^{t_1}  dt_2  \int_0^{t_2}  \bigg(\Big[H_1,[H_2,H_3]\Big]+\notag\\
    &\hspace{70pt}+\Big[[H_1,H_2],H_3\Big] \bigg)dt_3\notag\\
    &= \bigg(\frac{i}{6} \int_0^T  dt_1  \int_0^{t_1}  dt_2  \int_0^{t_2}  (u_3-2u_2+u_1)dt_3\bigg)\cdot\notag\\
    &\hspace{40pt}\cdot\Big[H_0,[H_0,H_c]\Big]+\notag\\
    &\hspace{20pt}+\bigg(\frac{i}{6} \int_0^T  dt_1  \int_0^{t_1}  dt_2  \int_0^{t_2}  (2u_1 u_3-u_1 u_2-\notag\\
    &\hspace{40pt}-u_2 u_3)dt_3\bigg)\Big[[H_c,H_0],H_c\Big]\notag\\
    &=d\Big[H_0,[H_0,H_c]\Big]+e\Big[[H_c,H_0],H_c\Big],
\end{align}
for scalars $d,e$. The rest of terms follow similarly.

For the Chebyshev approximation, we have

\begin{align}
    U(T)&\approx J_0(1)\mathbb{I}+2\sum_{i=1}^{p}J_i(1)T_i\notag\\
    &= (J_0(1)+2J_2(1))\mathbb{I}\notag\\
    &\hspace{50pt}+2J_1(1)\Omega^{(n)}+\notag\\
    &\hspace{60pt}+4J_2(1)(\Omega^{(n)})^2+\hdots\notag\\
    &=  s\mathbb{I}+l\Omega^{(n)}+q(\Omega^{(n)})^2+\hdots
\end{align}
for scalars $s,l,q$. Thus, we can obtain $U(T)$ only with the 3 MPO operations defined in the paper.

\section{Trotterization controlled Ising model}
\label{trotterization}

We consider a controlled Ising Hamiltonian

\begin{align}
H(t)&=H_Z^{odd}+H_Z^{even} + H_X(t)\\
&=\sum_{i \text{ odd}}^{N-1} J \sigma_i^z \sigma_{i+1}^z + \sum_{i \text{ even}}^{N-1} J \sigma_i^z \sigma_{i+1}^z + u(t)\sum_{i=1}^N\sigma_i^x.
\end{align}

And we write the total time evolution of the unitary operator via time discretisation as $U(T)=U_K U_{K-1}\hdots U_2 U_1$, where $U_j=e^{-i\Delta t H(\Delta t j)}$.

Now we let $U_j \approx e^{-i\Delta t H_X(\Delta t j)} e^{-i\Delta t H_Z^{odd}} e^{-i\Delta t H_Z^{even}}$, with an error of $O(\Delta t ^2)$, where each term can be factorised as

\begin{align}
    U_X^j = e^{-i\Delta t H_X(\Delta t j)}&=\prod_{i}^N e^{-i\Delta t u(\Delta t j) \sigma_i^x}\\
    U_Z^{\text{odd, } j} = e^{-i\Delta t H_Z^{odd}}&=\prod_{i \text{ odd}}^{N-1} e^{-i\Delta t J\sigma_i^z \sigma_{i+1}^z}\\
    U_Z^{\text{even, } j} = e^{-i\Delta t H_Z^{even}}&=\prod_{i \text{ even}}^{N-1} e^{-i\Delta t J\sigma_i^z \sigma_{i+1}^z}.
\end{align}

We can write these 3 exponentials as MPOs efficiently. For $U_X^j$ we have $1\times1\times2\times 2$ tensor cores 

\begin{equation}
    W^{[i]}_X= \begin{pmatrix}
        G(\phi)
    \end{pmatrix}, \hspace{5pt} G(\phi)=\begin{pmatrix}
        \cos(\phi) & -i\sin(\phi) \\
        -i\sin(\phi) & \cos(\phi)
    \end{pmatrix},
\end{equation}

where $\phi=u(\Delta t j)\Delta t$, for $i=1,\hdots, N$, so this MPO has bond dimension 1.  For the other two MPOs we first define the left and right tensor cores

\begin{equation}
    W_L = \begin{pmatrix}
        I & \sigma^z
    \end{pmatrix}, \hspace{5pt} W_R = \begin{pmatrix}
        \cos(\theta)I \\ -i\sin(\theta)\sigma^z
    \end{pmatrix},
\end{equation}
where $\theta = \Delta t J$. Now we let the tensor cores for $U_Z^{\text{odd, } j}$ be
\begin{equation}
    W_L, W_R, W_L, W_R,\hdots
\end{equation}
and if $N$ is odd we let the $N$th tensor core be $(I)$ instead of $W_L$. The tensor cores of $U_Z^{\text{even, } j}$ follow similarly
\begin{equation}
    (I), W_L, W_R, W_L,\hdots
\end{equation}

and if $N$ is even we let the $N$th tensor core be $(I)$ instead of $W_L$.

These MPOs have bond dimension of 1, 2 and 2 respectively, so the MPO representing the time step $U_j$ will have a bond dimension of $4$. And the bond dimension for the MPO representing the total time evolution for a time interval discretised into $K$ steps will be $4^K$.

\bibliography{main}

\end{document}